\documentclass[twocolumn]{article}

\usepackage{t1enc}
\usepackage[latin2]{inputenc}
\usepackage[english]{babel}

\usepackage{mathrsfs,amsmath}
\usepackage[scr=boondox]{mathalfa}
\usepackage{bm}

\usepackage{comment}
\usepackage{graphicx, float}
\usepackage{epstopdf}
\usepackage{cite}
\usepackage{url}
\usepackage{verbatim}
\usepackage[authoryear,round,longnamesfirst]{natbib}
\usepackage{cuted}

\usepackage{hyperref}
\hypersetup{
    bookmarks=true,         
    unicode=true,          
    pdftoolbar=true,        
    pdfmenubar=true,        
    pdffitwindow=false,     
    pdfstartview={FitH},    
    pdftitle={Rapidly rotating neutron stars with realistic nuclear matter equation of state},    
    pdfauthor={Balazs Kacskovics, Daniel Barta},     
    pdfsubject={Rotating Compact Stars},   
    pdfcreator={Kacskovics Balázs, Daniel Barta},   
    pdfproducer={Dr. Matyas Vasuth}, 
    pdfkeywords={compact stars} {numerical relativity} {LORENE}, 
    pdfnewwindow=true,      
    colorlinks=true,       
    linkcolor=red,          
    citecolor=green,        
    filecolor=magenta,      
    urlcolor=cyan           
}

\begin{document}

	\begin{titlepage}
		\title{Rapidly rotating neutron stars with realistic nuclear matter equation of state}
		\author{Bal\'{a}zs Kacskovics*, D\'{a}niel Barta and M\'{a}ty\'{a}s Vas\'{u}th}
	\end{titlepage}

	\maketitle
	
	\begin{strip}
    	\centering
    	\begin{minipage}{.8\textwidth}
        	\begin{abstract}
            	We performed a comparison of three different numerical codes for constructing equilibrium models; (I) a code for static equilibrium configurations, (II) an implementation of the Hartle--Thorne slow-rotation approximation, (III) a numerical solution of the full Einstein equations by \texttt{LORENE}. We aimed to construct sequences of uniformly rotating configurations at various rotation frequencies up to the Keplerian frequency for a hybrid hadronic--quark matter EOS where a smooth transition is provided between two separate phases. We investigated the difference of between the results computed by the implementation of Hartle--Thorne slow-rotation approximation and by \texttt{LORENE}/\texttt{nrotstar}, respectively. We have conclude that the codes can the difference between the slow-rotating and the fast-rotating approach increase exponentially, reaching $6.67\%$ for the maximal mass configuration rotating at the Keplerian frequency.
        	\end{abstract}
        	\hspace{2cm}
    	\end{minipage}
	\end{strip}
	






\section{Introduction}

In this paper we demonstrate that the structure of rotating compact stars can be described to a high level of accuracy at low frequencies by using the perturbative approach introduced by \cite{Hartle1967} and \cite{Hartle1968}, but the accuracy rapidly declines above $400$ Hz. The validity of the slow-rotation approximation is limited to angular frequencies $\omega^{2} \ll \omega_{\text{K}}$, where $\omega_{\text{K}}$ is the Keplerian angular frequency and terms only up to second order in $\omega$ are taken into account. By using a hybrid hadronic--quark EOS (that can already reach high gravitational mass in the static case), we focused in this study on the gravitational mass of numerically constructed  stellar solutions by both the perturbative and spectral methods of \texttt{LORENE}. In contrast \cite{Yagi2014} extracted the multipole moments (instead of the mass) by numerically constructing configurations with the \texttt{LORENE} and RNS codes (see the later in \cite{Stergioulas1995}).

\section{Realistic tabulated equation of state models} \label{sec:EoS}
In order to construct rotating stellar models that can already reach considerably high gravitational mass in the static case, we use a vector-meson extended linear $\sigma$-model that was recently published in \cite{Kovacs2022}. This zero-temperature EOS of finite quark (or baryon) chemical potential combines two relativistic mean-field (RMF) models, the softer SFHo model developed by \citet{Steiner2013} and the stiffer DD2 model of \citet{Typel2010,Hempel2010}. The major difference between the two hadronic models is that they have significant difference in their respective values for the slope of the symmetry energy $L$.

As temperature in the core of NSs is much less than the characteristic Fermi temperature, one can express the fluid's thermodynamic state at $T \simeq 0$ by the isentropic one-parameter equation of state (EOS)
\begin{equation}
	p = p(\epsilon)
\end{equation}
relating the isotropic pressure $p$ to the total energy density $\epsilon$. Interpolation is used over a the data set of energy and pressure values contained in a given EOS table to determine the function $p(\epsilon)$. The method of interpolation, however, has a decisive role for the accuracy of equilibrium models constructed with realistic (tabulated) EOS. Although global quantities are not affected significantly, the virial identities for realistic EOSs, are sensitive to what kind of interpolation is used, based on \cite{Nozawa1998}. In the procedures of \texttt{LORENE}, cubic Hermite splines are used for interpolation, where the Gibbs--Duhem relation can be used to replace the term $\nabla p/(\epsilon + p)$ by $\nabla H$ which appears in the hydrostationary equilibrium equation. On that account, instead of parametrizing the EOS by $\epsilon$, we shall use the log-enthalpy
\begin{equation} \label{enthalpy}
	H \equiv \ln \left(\frac{\epsilon+p}{c^{2}n_{B}}\right).
\end{equation}
Here $n_{B}$ is the the baryon-number density which is related to the baryon-mass density as $n_{B} = \rho_{B}/m_{u}$ and $m_{u} = 931.494$ MeV is the atomic mass unit, see in \cite{Barta2021}. The state variables such as the pressure, the energy density and the baryon-number density are directly obtained by
\begin{subequations} \label{EoS-eq}
	\begin{align}
		& \displaystyle \frac{\text{d}p}{\text{d}H} = \epsilon + p,   \\[10pt]
		& \displaystyle \epsilon = \frac{H}{p} \frac{\text{d}\log p}{\text{d}\log H},   \\[10pt]
		& n_{B} = \frac{\epsilon + p}{m_{B}c^{2}}\exp(-H).
	\end{align}
\end{subequations}
The advantage of using cubic Hermite splines for interpolation is that the Gibbs--Duhem relation is automatically satisfied in every point of the data set for the very reason that this interpolation reproduces not only the values of $\{p_{i},\, \epsilon_{i},\, (n_{B})_{i}\}$ themselves but also their respective derivatives, as shown by \cite{Nozawa1998}.

\section{Equations of the equilibrium stellar structure}
	The line element of a stationary, axisymmetric, and circular spacetime expressed in quasi-isotropic coordinates is
\begin{equation} \label{line-element}
	ds^{2} = -e^{2\tilde{\nu}}dt^{2} + e^{-2\tilde{\nu}}B^{2}r^2\sin^2\theta(d\phi - \omega dt)^{2} + e^{2(\zeta-\tilde{\nu})}(dr^{2} + r^{2}d\theta^{2}),
\end{equation}
where  $\tilde{\nu} = \tilde{\nu}(r,\theta)$, $\zeta = \zeta(r,\theta)$, $\omega = \omega(r,\theta)$, and $B = B(r,\theta)$ are metric functions, which depend on the coordinate radius $r$ and the angular coordinate $\theta$, as discussed by \cite{Stergioulas2003,Gourgoulhon2010}. In this spacetime, we consider equilibrium configurations of uniformly rotating stars with the 4-velocity 
\begin{equation} \label{fluid-velocity}
	u^{\mu} = (u^{0},\,0,\,0,\,\Omega u^{0})
\end{equation}
with $u^{0}$ determined by the normalization condition $u_{\mu}u^{\mu} = -1$ and they consist of a perfect fluid which is described by the stress--energy tensor, defined by
\begin{equation} \label{stress-energy tensor}
	T^{\alpha\beta} = (\epsilon + p)u^{\alpha}u^{\beta} + p g^{\alpha\beta}.
\end{equation}
In 3+1 formalism, that slices the four-dimensional spacetime by three-dimensional surfaces (\cite{Gourgoulhon2010}), and the Einstein field equations are reduced to the following linear partial differential equations (PDEs) of elliptic type for the gravitational fields $(\tilde{\nu},\, \zeta,\, \omega,\, B)$ expressed by \citet{Yagi2014}:
\begin{subequations} \label{Field-eq}
	\begin{align}
		& \displaystyle \Delta_{3}\tilde{\nu} = 4\pi e^{2(\zeta-\tilde{\nu})}\left[\left(2e^{2\tilde{\nu}}(u^{0})^{2}-1\right)(\rho + p) + 2p\right]   \nonumber \\
		& \displaystyle \hspace{2.8em} + \frac{1}{2}e^{-4\tilde{\nu}}B^{2}r^{2}\sin^{2}\theta\partial\omega\partial\omega - \partial\tilde{\nu}\partial(\ln B),   \label{Field-eq:a} \\[10pt]	
		& \displaystyle \tilde{\Delta}_{3}(\omega r\sin\theta) = 16\pi e^{2\zeta}(u^{0})^{2}(\rho + p)(\Omega-\omega)r\sin\theta   \nonumber \\
		& \displaystyle \hspace{6.2em} + r\sin\theta\partial\omega\partial(4\tilde{\nu} - 3\ln B),   \label{Field-eq:b} \\[10pt]
		& \displaystyle \Delta_{2}(B r\sin\theta) = 16\pi e^{2(\zeta-\tilde{\nu})}Bpr\sin\theta,   \label{Field-eq:c} \\[10pt]
		& \displaystyle \Delta_{2}\zeta = 8\pi e^{2(\zeta-\tilde{\nu})}\left[\left(2e^{2\tilde{\nu}}(u^{0})^{2}-1\right)(\rho + p) + p\right]   \nonumber \\
		& \displaystyle \hspace{2.8em} + \frac{3}{4}e^{-4\tilde{\nu}}B^{2}r^{2}\sin^{2}\theta\partial\omega\partial\omega - \partial\tilde{\nu} \partial\tilde{\nu},   \label{Field-eq:d}
	\end{align}
\end{subequations}
where the following second-order differential operators have been introduced:
\begin{subequations} \label{diffop-eq}
	\begin{align}
		& \displaystyle \Delta_{3} \equiv \frac{\partial^{2}}{\partial r^{2}} + \frac{2}{r}\frac{\partial}{\partial r} + \frac{1}{r^{2}}\frac{\partial^{2}}{\partial\theta^{2}} + \frac{\cos\theta}{r^{2}\sin\theta}\frac{\partial}{\partial \theta},   \\[10pt]
		& \displaystyle \tilde{\Delta}_{3} \equiv \Delta_{3} - \frac{1}{r^{2}\sin^{2}\theta},   \\[10pt]
		& \displaystyle \Delta_{2} = \frac{1}{r}\frac{\partial}{\partial r} + \frac{1}{r^{2}}\frac{\partial^{2}}{\partial\theta^{2}},   \\[10pt]
		& \displaystyle \partial f \partial g \equiv \frac{\partial f}{\partial r}\frac{\partial g}{\partial r} + \frac{1}{r^{2}}\frac{\partial f}{\partial \theta}\frac{\partial g}{\partial \theta}.
	\end{align}
\end{subequations}

As \cite{Gourgoulhon2010} pointed out; the operator $\Delta_{2}$ is a Laplacian in a 2-dimensional flat space spanned by the polar coordinates $(r,\theta)$, whereas $\Delta_{3}$ is the Laplacian in a 3-dimensional flat space, taking into account the axisymmetry $(\partial/\partial\varphi = 0)$. 

Furthermore, as \cite{Bonazzola1993} stated, under the assumption that the equilibrium configurations representing the gravitational field of this perfect fluid are stationary, axisymmetric and circular, the conservation of stress--energy turns into a simple first integral of motion:
\begin{equation} \label{first-integral-of-motion}
	H +\nu - \ln\Gamma = \text{const},
\end{equation}
in the form of an algebraic equation for the log-enthalpy \eqref{enthalpy}. The system of linear PDEs \eqref{diffop-eq} together with the first integral of fluid motion \eqref{first-integral-of-motion} is closed by the fluid's microscopic properties encoded into the EOS \eqref{EoS-eq} and completely determine the internal structure of rotating stars.

\section{Numerical Results}

	During our numerical analysis we compared the fast-rotating spectral code \href{https://lorene.obspm.fr}{\texttt{LORENE/rotstar}} made by LUTH group, an implematation of the slow-rotating Hartle--Thorne approach by \cite{Kovacs2022}, and a non-rotating model created by \cite{Barta2021} with each other.

	\subsection{LORENE/rotstar code}

		In \texttt{LORENE/rotstar} code we had three domain representing the \textit{nucleus}, the \textit{shell region} surrounding the nucleus and the \textit{compactified external region}. The elliptic equations are solved in each domain respectively, and matching conditions are imposed so that values of the metric functions and their derivatives agree on both sides of each domain, as it is discussed in detail in \citet{Yagi2014}. Functions used in \texttt{LORENE} are expanded in Chebyshev polynomials and trigonometric functions, respectively, and the latter are re-expanded in Legendre polynomials when it is advantageous. For that reason we distributed our grid points as follows: $(65, 33, 33, 17)$ grid points in $r$ and $4 \times 17$ grid points in $\theta$ in each one of the four domains to minimize the numerical error in our simulations.
		
		Further important parameters that was used in our numerical analysis are listed in table \ref{tab:intital_par}.
		
		\begin{table}[h!]
    		\centering
    		\begin{tabular}{rc|r}
        		$f_i$ & [Hz] & $0 \to 1350.030$ \\
        		$H_{0}$ & $[c^2]$ & $0.025 \to 0.460$ \\
        		$\Delta H$ & $[c^2]$ & $0.100$ \\
        		\hline
        		$\epsilon_{c}$ & [MeV/fm$^3$] & $74.400 \to 920.640$ \\        
        		$\Delta\epsilon$ & [MeV/fm$^3$] & $9.338$
    		\end{tabular} 
    		\caption{The range of important initial parameters that were used as input in the \texttt{LORENE/rotstar} code to construct stable equilibrium configurations. The last two rows represent the corresponding values of $H_{0}$ and $\Delta H$ expressed in energy densities.}\label{tab:intital_par}
\end{table}
	
	\subsection{Comparison}
		
\begin{figure}[h!]
			\begin{center}
				\includegraphics[width=0.41\textwidth]{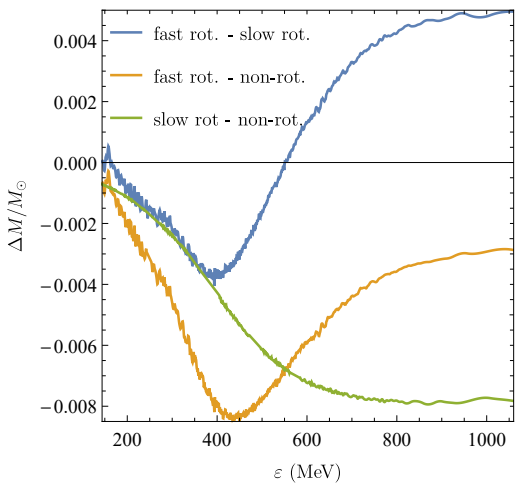}
				\caption{Here we show the difference of gravitational mass between the 3 algorithm we used at 0 Hz.} \label{fig:diff_mass_in_static}
			\end{center}
		\end{figure}

\begin{figure}[t!]
			\begin{center}
				\includegraphics[width=0.45\textwidth]{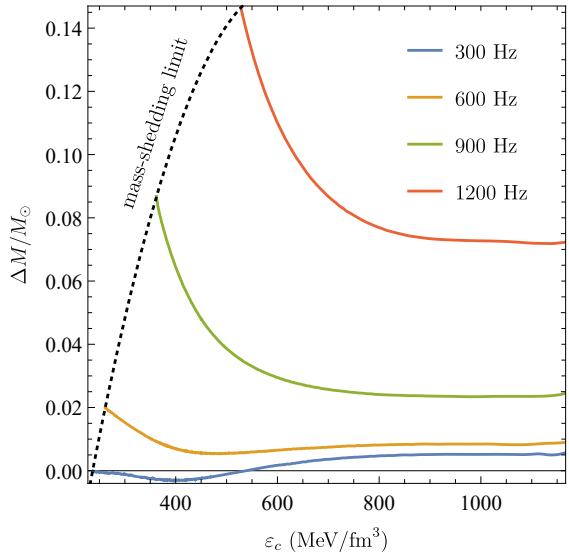}
				\caption{We show on this figure the difference of gravitational mass between the the fast- and slow-rotating approach at three different frequency.} \label{fig:diff_mass_rot}
			\end{center}
		\end{figure}

		First, we compared the three algorithm by measuring the absolute difference in gravitational mass computed by each code. Figure \ref{fig:diff_mass_in_static}displayes the absolute difference as a function of energy density for static configurations while figure \ref{fig:diff_mass_rot}does so for rotating ones. 
		
		We have found that both of our codes at low energy densities overestimates the mass compared to the one determined by \texttt{LORENE}. On higher energy densities the difference slightly decreases, and note that the slow-rotating approach starts to underestimate \texttt{LORENE}, which remains a characteristic feature of slow-rotating approach.
		
		Figure \ref{fig:diff_mass_rot} shows the difference of the gravitational mass evaluated by \texttt{LORENE} and the slow-rotating code for four different values of rotational frequency. The mass-shedding limit of rotating compact stars imposes a lower limit on the central energy density at each frequency. On low energy densities, we have found that near the limit the error of slow-rotating approach increases exponentially as the frequency approaches the Keplerian limit. At $300$ Hz, as a remnant, the overestimation of the static case at low-energy density is still visible. This underestimation is gradually decreasing and converges to a constant value towards higher energy densities.
	
\begin{figure}[h!]
			\begin{center}
				\includegraphics[width=0.45\textwidth]{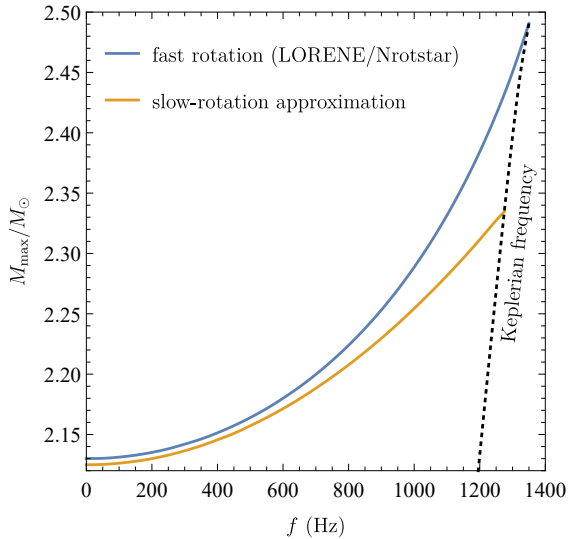}
				\caption{This figure shows the maximum gravitational mass reached with the fast- and slow-rotating approach as the function of the frequency. Furthermore, the black dashed line denotes the mass-shedding limit, which contains all the configurations at the Keplerian frequency.} \label{fig:mass_diff_freq}
			\end{center}
		\end{figure}

		Also confirmed by figure \ref{fig:mass_diff_freq}, where we have displayed the gravitational mass as the function of frequency. As we closing on the mass-shedding limit the difference between the slow-rotating and the fast-rotating approach grows significantly. To be exact, the discrepancy between the two method is $6.67\%$, and maximum masses at the mass-shedding limit respectively is $2.49~\text{M}_\odot$ for \texttt{LORENE} and $2.34~\text{M}_\odot$ for Hartle--Thorne method. The frequency, the central energy density and equatorial radius of the maximum configuration evaluated by the two method is the following:  $f^\text{L} = 1350.03$ Hz and $f^\text{HT} = 1275.33$ Hz, $\epsilon^\text{L}_\text{c}= 920.64$ MeV/fm$^3$ and $\epsilon^\text{HT}_\text{c} = 888.39$ MeV/fm$^3$, $R^\text{L}_\text{eq} = 12.96$ km and $R^\text{HT}_\text{eq} = 12.92$ km. 
	
\begin{figure}[h!]
			\begin{center}
				\includegraphics[width=0.45\textwidth]{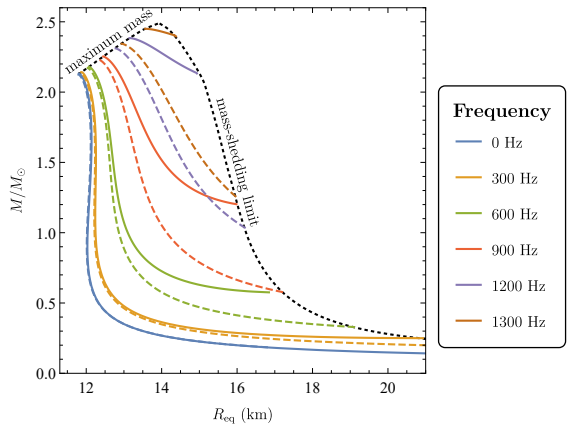}
				\caption{The gravitational mass--equatorial radius relations for static and rotating compact stars with a combined EOS of SFHo/DD2 models inside their core. The solid lines represent sequences computed by LORENE and dashed lines represents those of our slow-rotating model on different frequencies.}\label{fig:mr_plot}
			\end{center}
		\end{figure}

		The figure \ref{fig:mr_plot}shows the gravitational mass $M$ vs. the gauge-invariant circumferential stellar radius at the equator $R_{\text{eq}}$. The circumferential equatorial radius is defined as
		\begin{equation}
			R_{\text{eq}} = e^{-\tilde{\nu}}Br_{\text{eq}},
		\end{equation}
		where $r_{\text{eq}}$ is the coordinate equatorial radius in quasi-isotropic coordinates where $p = 0$ at $\theta = \pi/2$. 
		
		The mass--radius relations are shown for two sets of curves: solid lines designate sequences of equilibrium configurations which were computed by the spectral method in the \texttt{LORENE}/\texttt{nrotstar} codes, and dashed lines (of the same colour) designate those that were produced by the Hartle--Thorne slow-rotation approximation. The blue line denotes the static (non-rotating) sequence, while the yellow, green, red, purple and brown lines denote sequences of configurations rotating with constant spin frequencies of 300 Hz, 600 Hz, 900 Hz, 1200 Hz and 1300 Hz, respectively. The black-coloured dotted lines marked with inscriptions represent the maximum mass configurations and mass-shedding configurations, respectively, and thus, indicate the limit where sufficient conditions are met for the instability with respect to the axisymmetric perturbations.
		
		The mass--radius relations exhibit the features that are in accordance with our previous findings. Regardless of frequency, the mass computed by slow-rotating method is lower than in the fast-rotating case. Furthermore, this difference of gravitational mass in the two model is decreasing with the increasing central energy density.

\section{Conclusions}

Our aim in this paper to show that together with a hybrid hadronic--quark matter EOS that was interpolated smoothly result in crossover phase transitions by \citet{Kovacs2022}, and with rotation higher masses can be reached. For that reason, we used two different method in this paper, a slow- and a fast-rotating method. Because the Hartle--Thorne slow-rotation approximation is well-known, here we restricted our discussion on to the introduction of the equations that determine the equilibrium configurations of rotating stars in \texttt{LORENE}, which has developed by \citet{LUTH1997}.

We demonstrated that the slow-rotating approximation -- apart from very low energy densities and frequencies -- underestimates the results of corresponding configurations computed by \texttt{LORENE}. This discrepancy decreases with the growing energy density, but its exponentially increases with the frequency. The configuration with maximum mass of $2.49~M_\odot$ for \texttt{LORENE} and $2.34~M_\odot$ for slow-rotating method at Keplerian frequency of $1350.03$ Hz and $1275.33$ Hz, respectively. Through this analysis of discrepancy in these computations proved the accuracy of \texttt{LORENE} in high-frequency regime.

\section{Acknowledgments}
The authors wish to acknowledge the generous hospitality of the LUTH departent at the Observatoire de Paris--PSL, Meudon, where part of this paper was completed. We would like to especially thank J. Novak and E. Gourgoulhon for the constructive and very helpful discussions on \texttt{LORENE}. We would also like to extend our gratitude to P. Kov\'{a}cs, J. Tak\'{a}tsy, Gy. Wolf for providing us with the tabulated EOS of the applied vector-meson model. Last but not least, we would like to thank the WSC GPU Laboratory at the Wigner RCP for providing us with access to high-end computational resources. This research has been supported in part by the the National Research, Development and Innovation Office (Hungarian abbreviation: NKFIH) under OTKA grant agreement No. K138277.

\bibliographystyle{plainnat}
\bibliography{artbib_iwara2022.bib}

\begin{thebibliography}{14}
\providecommand{\natexlab}[1]{#1}
\providecommand{\url}[1]{\texttt{#1}}
\expandafter\ifx\csname urlstyle\endcsname\relax
  \providecommand{\doi}[1]{doi: #1}\else
  \providecommand{\doi}{doi: \begingroup \urlstyle{rm}\Url}\fi

\bibitem[Barta(2021)]{Barta2021}
D\'{a}niel Barta.
\newblock Fundamental and higher-order excited modes of radial oscillation of
  neutron stars for various types of cold nucleonic and hyperonic matter.
\newblock \emph{Class. Quant. Grav.}, 38\penalty0 (18):\penalty0 185002, 09
  2021.
\newblock \doi{10.1088/1361-6382/ac12e2}.

\bibitem[{Bonazzola} et~al.(1993){Bonazzola}, {Gourgoulhon}, {Salgado}, and
  {Marck}]{Bonazzola1993}
S.~{Bonazzola}, E.~{Gourgoulhon}, M.~{Salgado}, and J.~A. {Marck}.
\newblock {Axisymmetric rotating relativistic bodies: A new numerical approach
  for 'exact' solutions}.
\newblock \emph{Astron. Astrophy.}, 278\penalty0 (2):\penalty0 421--443,
  November 1993.

\bibitem[{Gourgoulhon}(2010)]{Gourgoulhon2010}
Eric {Gourgoulhon}.
\newblock An introduction to the theory of rotating relativistic stars, mar
  2010.

\bibitem[Gourgoulhon et~al.(1997)Gourgoulhon, Grandclement, Marck, Novak, and
  Taniguchi]{LUTH1997}
Eric Gourgoulhon, Philippe Grandclement, Jean-Alain Marck, Jerome Novak, and
  Keisuke Taniguchi.
\newblock Lorene.
\newblock 1997.
\newblock URL \url{https://lorene.obspm.fr/}.

\bibitem[{Hartle}(1967)]{Hartle1967}
James~B. {Hartle}.
\newblock {Slowly Rotating Relativistic Stars. I. Equations of Structure}.
\newblock \emph{\apj}, 150:\penalty0 1005, December 1967.
\newblock \doi{10.1086/149400}.

\bibitem[{Hartle} and {Thorne}(1968)]{Hartle1968}
James~B. {Hartle} and Kip~S. {Thorne}.
\newblock {Slowly Rotating Relativistic Stars. II. Models for Neutron Stars and
  Supermassive Stars}.
\newblock \emph{Astrophys. J.}, 153:\penalty0 807, September 1968.
\newblock \doi{10.1086/149707}.

\bibitem[{Hempel} and {Schaffner-Bielich}(2010)]{Hempel2010}
Matthias {Hempel} and J\"urgen {Schaffner-Bielich}.
\newblock A statistical model for a complete supernova equation of state.
\newblock \emph{Nuclear Physics A}, 837\penalty0 (3):\penalty0 210--254, 2010.
\newblock ISSN 0375-9474.
\newblock \doi{https://doi.org/10.1016/j.nuclphysa.2010.02.010}.

\bibitem[{Kov\'acs} et~al.(2022){Kov\'acs}, {Tak\'atsy}, {Schaffner-Bielich},
  and {Wolf}]{Kovacs2022}
P\'eter {Kov\'acs}, J\'anos {Tak\'atsy}, J\"urgen {Schaffner-Bielich}, and
  Gy\"orgy {Wolf}.
\newblock Neutron star properties with careful parametrization in the vector
  and axial-vector meson extended linear sigma model.
\newblock \emph{Phys. Rev. D}, 105:\penalty0 103014, May 2022.
\newblock \doi{10.1103/PhysRevD.105.103014}.

\bibitem[{Nozawa} et~al.(1998){Nozawa}, {Stergioulas}, {Gourgoulhon}, and
  {Eriguchi}]{Nozawa1998}
Tetsuo {Nozawa}, Nikolaos {Stergioulas}, Eric {Gourgoulhon}, and Yoshiharu
  {Eriguchi}.
\newblock Construction of highly accurate models of rotating neutron stars
  comparison of three different numerical schemes.
\newblock \emph{Astron. Astrophys. Suppl. Ser.}, 132\penalty0 (3):\penalty0
  431--454, 1998.
\newblock \doi{10.1051/aas:1998304}.

\bibitem[{Steiner} et~al.(2013){Steiner}, {Hempel}, and {Fischer}]{Steiner2013}
Andrew~W. {Steiner}, Matthias {Hempel}, and Tobias {Fischer}.
\newblock Core-collapse supernova equations of state based on neutron star
  observations.
\newblock \emph{The Astrophysical Journal}, 774\penalty0 (1):\penalty0 17, aug
  2013.
\newblock \doi{10.1088/0004-637X/774/1/17}.

\bibitem[Stergioulas(2003)]{Stergioulas2003}
Nikolaos Stergioulas.
\newblock Rotating stars in relativity.
\newblock \emph{Living Reviews in Relativity}, 6:\penalty0 3, 06 2003.
\newblock \doi{10.12942/lrr-2003-3}.

\bibitem[{Stergioulas} and {Friedman}(1995)]{Stergioulas1995}
Nikolaos {Stergioulas} and John~L. {Friedman}.
\newblock {Comparing Models of Rapidly Rotating Relativistic Stars Constructed
  by Two Numerical Methods}.
\newblock \emph{The Astrophysical Journal}, 444:\penalty0 306, May 1995.
\newblock \doi{10.1086/175605}.

\bibitem[Typel et~al.(2010)Typel, R\"opke, Kl\"ahn, Blaschke, and
  Wolter]{Typel2010}
S.~Typel, G.~R\"opke, T.~Kl\"ahn, D.~Blaschke, and H.~H. Wolter.
\newblock Composition and thermodynamics of nuclear matter with light clusters.
\newblock \emph{Phys. Rev. C}, 81:\penalty0 015803, Jan 2010.
\newblock \doi{10.1103/PhysRevC.81.015803}.

\bibitem[Yagi et~al.(2014)Yagi, Kyutoku, Pappas, Yunes, and
  Apostolatos]{Yagi2014}
Kent Yagi, Koutarou Kyutoku, George Pappas, Nicol{\'a}s Yunes, and
  Theocharis~A. Apostolatos.
\newblock Effective no-hair relations for neutron stars and quark stars:
  Relativistic results.
\newblock \emph{\prd}, 89\penalty0 (12):\penalty0 124013, June 2014.
\newblock \doi{10.1103/PhysRevD.89.124013}.

\end{thebibliography}

\end{document}